\documentclass[twoside,slac_one]{revtex4}
\usepackage{graphicx}
\usepackage{fancyhdr}
\usepackage{amsmath} 
\usepackage{bm}
\usepackage{amsxtra}
\usepackage{amssymb}
\usepackage{amsthm}
\usepackage{latexsym}
\usepackage{lscape}
\usepackage{cancel}

\begin{document}

\title{Recent Results of Fluctuation and Correlation Studies from the QCD Critical Point Search at RHIC}

%

\author{T.~Tarnowsky}
\affiliation{National Superconducting Cyclotron Laboratory, Michigan State University, East Lansing, MI, USA}

\begin{abstract}
Enhanced fluctuations and correlations have been observed in the phase transitions of many systems. Their appearance at the predicted QCD phase transition (especially near the expected critical point) may provide insight into the nature of the phase transition. Recent results from the QCD Critical Point Search at RHIC will be presented, with a focus on particle ratio ($K/\pi$, $p/\pi$, and $K/p$) fluctuations and their comparison to previous measurements and theoretical predictions.
\end{abstract}

\maketitle

\thispagestyle{fancy}


\section{Introduction}
Fluctuations and correlations are well known signatures of phase transitions. In particular, the quark/gluon to hadronic phase transition may lead to significant fluctuations \cite{Koch1}. In 2010, the Relativistic Heavy Ion Collider (RHIC) began a program to search for the QCD critical point. This involves an ``energy scan'' of Au+Au collisions from top collision energy ($\sqrt{s_{NN}}$ = 200 GeV) down to energies as low as $\sqrt{s_{NN}}$ = 7.7 GeV \cite{STARBES}. This critical point search will make use of the study of correlations and fluctuations, particularly those that could be enhanced during a phase transition that passes close to the critical point. Particle ratio fluctuations are an observable that has already been studied as a function of energy and system-size. Combining these previous measurements with new results from the energy scan will provide additional information to assist in locating the QCD critical point.
\\

One of the suggested signatures for formation of a deconfined phase of quark-gluon matter was the enhancement of strangeness production in nucleus-nucleus collisions compared to hadron-hadron collisions \cite{Rafelski1}. One mechanism to produce this strangeness enhancement in a nucleus-nucleus collision that forms a partonic system is from pair production of $s\overline{s}$ via gluon fusion. Enhancement of particles containing strange particles (relative to production in p+Be and p+Pb interactions) was observed in Pb+Pb collisions at the CERN SPS \cite{NA57}. Particles containing larger numbers of strange/anti-strange quarks (e.g. $\Omega^{0}$) were seen to be more enhanced than particles containing fewer strange/anti-strange quarks (e.g. $\Lambda^{0}$).
\\

Studies of the ratio of strange hadrons to pions by the NA49 experiment at the CERN SPS found a maximum in the $K^{+}/\pi^{+}$ ratio as a function of energy at intermediate SPS energies ($\sqrt{s_{NN}}$ = 7.6 GeV) \cite {NA49_KPi}. Further measurements at higher SPS energies (up to $\sqrt{s_{NN}}$ = 17.3 GeV) and RHIC energies (up to $\sqrt{s_{NN}}$ = 200 GeV) indicated a plateau in the $K^{+}/\pi^{+}$ ratio may be reached at high energies \cite{NA49_KPi,STAR_KPi,STAR_KPi2}. With measurements of the mean $K/\pi$ ratio (first moment of the $K/\pi$ distribution) as a function of energy demonstrating a potential maximum, a logical next step was the study of fluctuations in the average $K/\pi$ ratio (second moment of the $K/\pi$ distribution).
\\

Dynamical particle ratio fluctuations, specifically fluctuations in the $K/\pi$, $p/\pi$, and $K/p$ ratio, can provide information on the quark-gluon to hadron phase transition \cite{Strangeness1, Strangeness2, Strangeness3}. The variable used to measure these fluctuations is referred to as $\nu_{dyn}$. $\nu_{dyn}$ was originally introduced to study net charge fluctuations \cite{nudyn1, nudyn2}. $\nu_{dyn}$ quantifies deviations in the particle ratios from those expected for an ideal statistical Poissonian distribution. The definition of $\nu_{dyn,K/\pi}$ (describing fluctuations in the $K/\pi$ ratio) is,
\begin{eqnarray}
\nu_{dyn,K/\pi} = \frac{\left<N_{K}(N_{K}-1)\right>}{\left<N_{K}\right>^{2}}
+ \frac{\left<N_{\pi}(N_{\pi}-1)\right>}{\left<N_{\pi}\right>^{2}}
- 2\frac{\left<N_{K}N_{\pi}\right>}{\left<N_{K}\right>\left<N_{\pi}\right>}\ ,
\label{nudyn}
\end{eqnarray}
where $N_{K}$ and $N_{\pi}$ are the number of kaons and pions in a particular event, respectively. In this proceeding, $N_{K}$ and $N_{\pi}$ are the total charged multiplicity for each particle species. A formula similar to (\ref{nudyn}) can be constructed for $p/\pi$ and other particle ratio fluctuations. By definition, $\nu_{dyn}$ = 0 for the case of a Poisson distribution of kaons and pions. It is also largely independent of detector acceptance and efficiency in the region of phase space being considered \cite{nudyn2}. An in-depth study of $K/\pi$ fluctuations in Au+Au collisions at $\sqrt{s_{NN}}$ = 200 and 62.4 GeV was previously carried out by the STAR experiment \cite{starkpiprl}. 
\\

Comprehensive studies of efficiency and acceptance effects on $\nu_{dyn}$ have been shown in \cite{Christiansen,ehaslum}. Using simulations, those studies demonstrate that $\nu_{dyn,K/\pi}$ is independent of particle detection efficiency and independent of azimuthal acceptance for pairs of particles that are randomly distributed, but not for the extreme cases where all particle pairs are produced aligned or back-to-back. Also shown are the effects of particle misidentification and double counting of tracks. Those simulations indicate that misidentifying pions as kaons leads to a decrease in $\nu_{dyn}$ as the level of misidentification grows.
\\

Earlier measurements of particle ratio fluctuations utilized the variable $\sigma_{dyn}$ \cite{NA49_kpi_ppi}, 
\begin{equation}
\sigma_{dyn} = sgn(\sigma_{data}^{2}-\sigma_{mixed}^{2})\sqrt{|\sigma_{data}^{2}-\sigma_{mixed}^{2}|}\ ,
\label{signudyn}
\end{equation}
where $\sigma$ is the relative width of the $K/\pi$ (or $p/\pi$, or $K/p$) distribution in either real data or mixed events. It has been shown that $\nu_{dyn}$ is a first order expansion of $\sigma_{dyn}$ about the mean of its denominator \cite{jeon,baym,sdasthesis}. The two variables are related as $\sigma_{dyn}^{2} = \nu_{dyn}$. Their relationship is derived below for $K/\pi$ fluctuations following the arguments in \cite{baym,sdasthesis}, but the derivation can be generalized for $p/\pi$ and $K/p$ fluctuations.

The relative fluctuation of a quantity, $x$, can be expressed as,

\begin{equation}
\sigma_{x} = \frac{\sqrt{\left<x^{2}\right>-\left<x\right>^{2}}}{\left<x\right>}\ ,
\label{SDev}
\end{equation}

and the relative variance of $x$ is then,

\begin{equation}
\sigma^{2}_{x} = \frac{<x^{2}>-<x>^{2}}{<x>^{2}}\ .
\label{Variance}
\end{equation}

Equation \ref{Variance} is the relevant quantity measured in \cite{NA49_kpi_ppi,NA49_kp} from the data and mixed events to construct $\sigma_{dyn}$.

Equation 12 from \cite{baym} expresses the fluctuations in the average $K/\pi$ ratio ($\left<K/\pi\right>$) as,

%
\begin{equation}
D^{2} \equiv \frac{\omega_{K/\pi}}{\left<K/\pi\right>} = \frac{\omega_{K}}{\left<K\right>}+\frac{\omega_{\pi}}{\left<\pi\right>}-2\frac{\left<K\pi\right>-\left<K\right>\left<\pi\right>}{\left<K\right>\left<\pi\right>}\ ,
\label{Dispersion}
\end{equation}

with $\omega_{x} = \frac{\left<x^{2}\right>-\left<x\right>^{2}}{\left<x\right>}$.

Then it can be seen from Equations \ref{SDev} and \ref{Dispersion}, and the definition of $\omega$, that $D^{2} \equiv \sigma^{2}$. Following from Equation \ref{Dispersion}, $\sigma^{2}_{K/\pi}$ can be written as,

\begin{equation}
\sigma^{2}_{K/\pi} = \frac{\left<N_{K}^{2}\right>-\left<N_{K}\right>^{2}}{\left<N_{K}\right>^{2}}+\frac{\left<N_{\pi}^{2}\right>-\left<N_{\pi}\right>^{2}}{\left<N_{\pi}\right>^{2}}-2\frac{\left<N_{K}N_{\pi}\right>-\left<N_{K}\right>\left<N_{\pi}\right>}{\left<N_{K}\right>\left<N_{\pi}\right>}\ .
\label{sigmakpi}
\end{equation}

The first term in Equation \ref{sigmakpi} can be expanded, yielding,

\begin{equation}
\frac{\left<N_{K}^{2}\right>-\left<N_{K}\right>^{2}}{\left<N_{K}\right>^{2}} = \frac{\left<N_{K}\left(N_{K}-1\right)\right>-\left<N_{K}\right>\left(\left<N_{K}\right>-1\right)}{\left<N_{K}\right>} = \frac{\left<N_{K}\left(N_{K}-1\right)\right>}{\left<N_{K}\right>^{2}}+\frac{1}{\left<N_{K}\right>}-1\ ,
\label{firstterm}
\end{equation}

and similarly for the second term of Equation \ref{sigmakpi},

\begin{equation}
\frac{\left<N_{\pi}^{2}\right>-\left<N_{\pi}\right>^{2}}{\left<N_{\pi}\right>^{2}} = \frac{\left<N_{\pi}\left(N_{\pi}-1\right)\right>-\left<N_{\pi}\right>\left(\left<N_{\pi}\right>-1\right)}{\left<N_{\pi}\right>} = \frac{\left<N_{\pi}\left(N_{\pi}-1\right)\right>}{\left<N_{\pi}\right>^{2}}+\frac{1}{\left<N_{\pi}\right>}-1\ .
\label{secondterm}
\end{equation}

Substituting Equations \ref{firstterm} and \ref{secondterm} into Equation \ref{sigmakpi},

\begin{eqnarray}
\sigma^{2}_{K/\pi} = \frac{\left<N_{K}\left(N_{K}-1\right)\right>}{\left<N_{K}\right>^{2}}+\frac{1}{\left<N_{K}\right>}-1+\frac{\left<N_{\pi}\left(N_{\pi}-1\right)\right>}{\left<N_{\pi}\right>^{2}}+\frac{1}{\left<N_{\pi}\right>}-1-2\frac{\left<N_{K}N_{\pi}\right>-\left<N_{K}\right>\left<N_{\pi}\right>}{\left<N_{K}\right>\left<N_{\pi}\right>}\\
= \frac{\left<N_{K}\left(N_{K}-1\right)\right>}{\left<N_{K}\right>^{2}}+\frac{1}{\left<N_{K}\right>}-\cancel{1}+\frac{\left<N_{\pi}\left(N_{\pi}-1\right)\right>}{\left<N_{\pi}\right>^{2}}+\frac{1}{\left<N_{\pi}\right>}-\cancel{1}-2\frac{\left<N_{K}N_{\pi}\right>}{\left<N_{K}\right>\left<N_{\pi}\right>}+\cancel{2}\ , \notag
\end{eqnarray}

and after ordering terms,

\begin{equation}
\sigma^{2}_{K/\pi} = \frac{\left<N_{K}\left(N_{K}-1\right)\right>}{\left<N_{K}\right>^{2}}+\frac{\left<N_{\pi}\left(N_{\pi}-1\right)\right>}{\left<N_{\pi}\right>^{2}}-2\frac{\left<N_{K}N_{\pi}\right>}{\left<N_{K}\right>\left<N_{\pi}\right>}+\frac{1}{\left<N_{K}\right>}+\frac{1}{\left<N_{\pi}\right>}\ .
\label{sigma_nu_compare}
\end{equation}

Comparing to Equation \ref{nudyn}, Equation \ref{sigma_nu_compare} can be expressed as,

\begin{equation}
\sigma^{2}_{K/\pi} = \nu_{dyn,K/\pi}+\frac{1}{\left<N_{K}\right>}+\frac{1}{\left<N_{\pi}\right>}\ ,
\end{equation}

and from \cite{nudyn1}, $\nu_{stat}$ would be defined as,

\begin{equation}
\nu_{stat,K/\pi} = \frac{1}{\left<N_{K}\right>}+\frac{1}{\left<N_{\pi}\right>}\ ,
\end{equation}

so finally,

\begin{equation}
\sigma^{2}_{K/\pi} = \nu_{dyn,K/\pi}+\nu_{stat,K/\pi}\ ,
\end{equation}

and,

\begin{eqnarray}
\nu_{dyn,K/\pi} = \sigma^{2}_{K/\pi} - \nu_{stat,K/\pi} \\
= \sigma^{2}_{K/\pi} - \sigma_{stat,K/\pi}^{2} 
\end{eqnarray}

\begin{equation}
\nu_{dyn,K/\pi} = \sigma^{2}_{dyn,K/\pi}\ .
\label{nu_sigma_equal}
\end{equation}

As mentioned earlier, Equation \ref{nu_sigma_equal} can be generalized for other particle ratio species.

\section{Experimental Analysis}
The data presented here for $K/\pi$, $p/\pi$, and $K/p$ fluctuations was acquired by the STAR experiment at RHIC from minimum bias (MB) Au+Au collisions at center-of-mass collision energies ($\sqrt{s_{NN}}$) of 200, 62.4, 39, 11.5, and 7.7 GeV \cite{STAR}. The main particle tracking detector at STAR is the Time Projection Chamber (TPC) \cite{STARTPC}. All detected charged particles in the pseudorapidity interval $|\eta| < 1.0$ were measured. The transverse momentum ($p_{T}$) range for pions and kaons was $0.2 < p_{T} < 0.6$ GeV/$c$, and for protons was $0.4 < p_{T} < 1.0$ GeV/$c$. Charged particle identification involved measured ionization energy loss ($dE/dx$) in the TPC gas and total momentum ($p$) of the track. The energy loss of the identified particle was required to be less than two standard deviations (2$\sigma$) from the predicted energy loss of that particle at a particular momentum. A second requirement was that the measured energy loss of a pion/kaon was more than 2$\sigma$ from the energy loss prediction of a kaon/pion, and similarly for $p/\pi$ and $K/p$ measurements.

Data was also analyzed using the recently completed Time of Flight (TOF) detector \cite{STARTOF}. The TOF is a multi-gap resistive plate chamber (MRPC) detector. Particle identification was carried out using the time-of-flight information for a track along with its momentum, determined by the TPC. From these two quantities the particle's mass can be calculated. Utilizing the TOF allows the identified particle momentum reach to be extended. TOF identification in these measurements extends the momentum range of pions and kaons up to $1.4$ GeV/$c$ and protons up to $1.8$ GeV/$c$.

\section{Results and Discussion}

\begin{figure}
\includegraphics[width=0.74\textwidth]{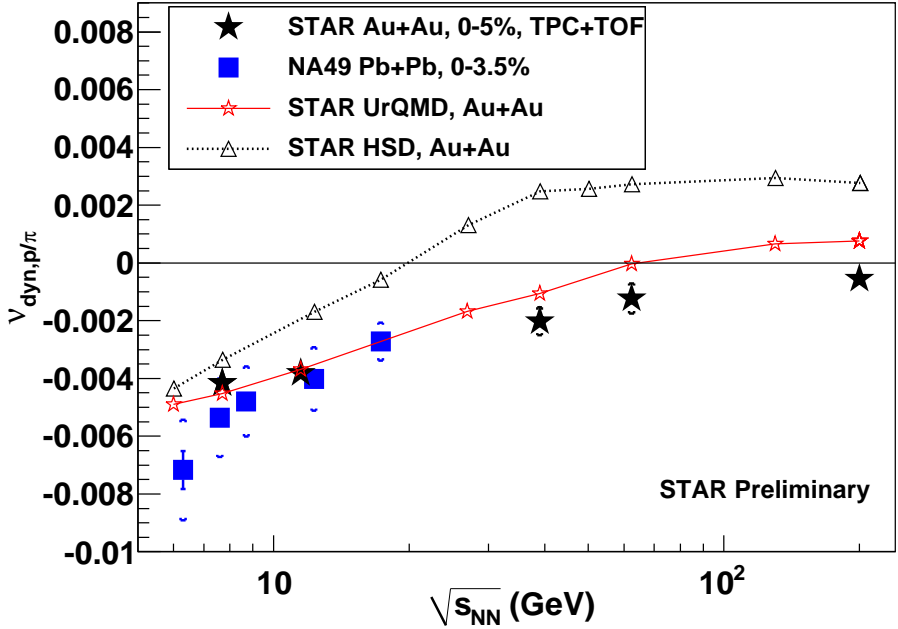}
\caption{Results for the measurement of $\nu_{dyn,p/\pi}$ as measured by the STAR TPC+TOF (black stars) from central 0-5\% Au+Au collisions at $\sqrt{s_{NN}}$ = 7.7-200 GeV.  Also shown are results from the NA49 (solid blue squares) from central 0-3.5\% Pb+Pb collisions. Model predictions from UrQMD and HSD using the STAR experimental acceptance (red and black lines, respectively) are also included.}
\label{ppi_excitation}
\end{figure}

\begin{figure}
\includegraphics[width=0.75\textwidth]{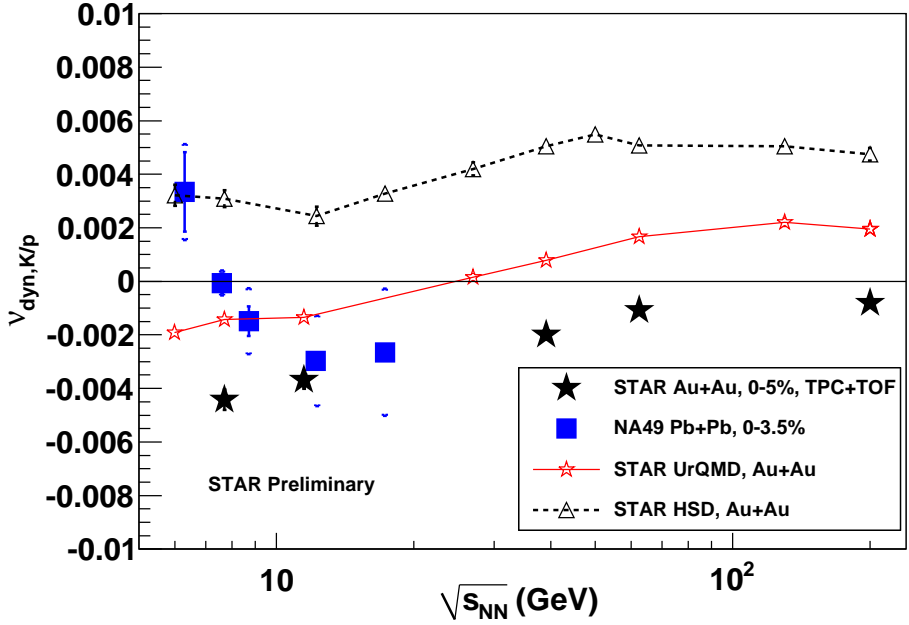}
\caption{Results for the measurement of $\nu_{dyn,K/p}$ as measured by the STAR TPC+TOF (black stars) from central 0-5\% Au+Au collisions at $\sqrt{s_{NN}}$ = 7.7-200 GeV.  Also shown are results from the NA49 (solid blue squares) from central 0-3.5\% Pb+Pb collisions. Model predictions from UrQMD and HSD using the STAR experimental acceptance (red and black lines, respectively) are also included.}
\label{kp_excitation}
\end{figure}

\begin{figure}[]
\includegraphics[width=0.75\textwidth]{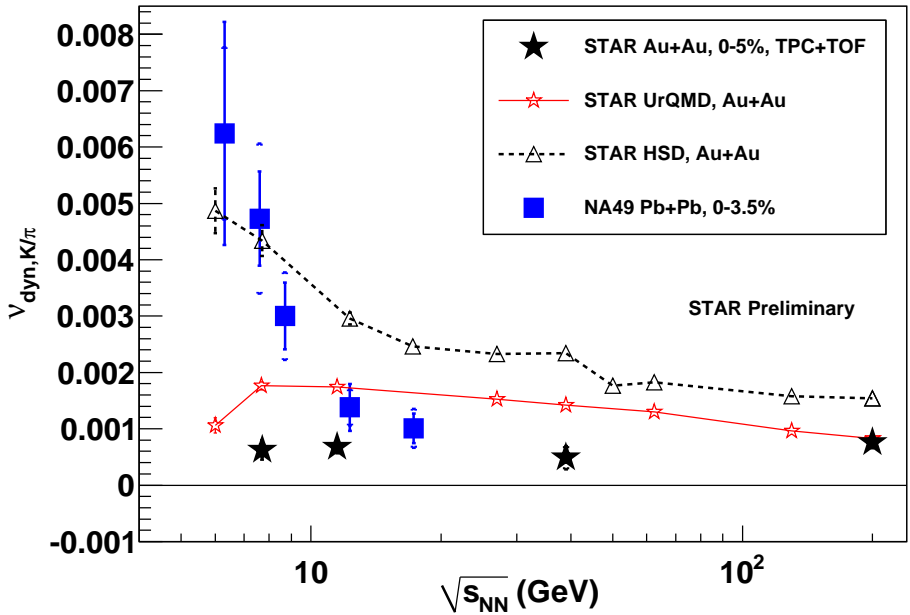}
\caption{Results for the measurement of $\nu_{dyn,K/\pi}$ (left) as measured by the STAR TPC+TOF (black stars) from central 0-5\% Au+Au collisions at $\sqrt{s_{NN}}$ = 7.7-200 GeV. Also shown are the same measurements from the NA49 (solid blue squares) from central 0-3.5\% Pb+Pb collisions. Model predictions from UrQMD and HSD using the STAR experimental acceptance (red and black lines, respectively) are also included.}
\label{kpi_excitation}
\end{figure}

Results for dynamical particle ratio fluctuations from the RHIC energy scan are shown in Figures \ref{ppi_excitation}, \ref{kp_excitation}, and \ref{kpi_excitation}. Figure \ref{ppi_excitation} shows the measured dynamical $p/\pi$ fluctuations as a function of incident energy, expressed as $\nu_{dyn,p/\pi}$. The measured fluctuations from the STAR experiment are plotted as the black stars, from the NA49 experiment \cite{NA49_kpi_ppi} at the SPS as blue squares, and two transport model predictions from UrQMD and HSD (red and black lines, respectively), using the STAR experimental acceptance. All model predictions in this proceeding are processed through the same acceptance. The published NA49 results have been converted to $\nu_{dyn}$ using the relation $\sigma_{dyn}^{2} \approx \nu_{dyn}$. STAR measures a general trend from larger, negative values of $p/\pi$ fluctuations at lower energies, which then increases towards zero at $\sqrt{s_{NN}}$ = 200 GeV. These fluctuations are negative due to the dominance of the third term (correlated production, e.g. $\Delta^{++} \rightarrow p^{+} + \pi^{+}$) in $\nu_{dyn,p/\pi}$. There is overlap between the two different experiments at the lowest energies measured by STAR ($\sqrt{s_{NN}}$ = 7.7 and 11.5 GeV), but the slopes of the excitation functions from both experiments appear to be different. It is possible that in the future RHIC will collide Au+Au at an energy below $\sqrt{s_{NN}}$ = 7.7 GeV, providing data point(s) to test the trends observed in the first part of the search for the QCD critical point. \\

Figure \ref{kp_excitation} shows the measured dynamical $K/p$ fluctuations as a function of incident energy, expressed as $\nu_{dyn,K/p}$. Results from the NA49 experiment are from \cite{NA49_kp}. 
Dynamical $K/p$ fluctuations show an increase from larger, negative values toward zero as a function of increasing incident energy. This is qualitatively and quantitatively similar to the trend measured for dynamical $p/\pi$ fluctuations. 
The experimental trend observed by STAR at the lowest incident energies ($\sqrt{s_{NN}}$ = 7.7 and 11.5 GeV) is dramatically different than that measured by NA49. The STAR result is consistent with negative dynamical $K/p$ fluctuations down to $\sqrt{s_{NN}}$ = 7.7 GeV, while that from NA49 is consistent with zero and increases to a positive value at $\sqrt{s_{NN}}$ = 6.3 GeV.\\ 

Figure \ref{kpi_excitation} shows the measured dynamical $K/\pi$ fluctuations as a function of incident energy, expressed as $\nu_{dyn,K/\pi}$. 
Results from the NA49 experiment are from \cite{NA49_kpi_ppi}. STAR measures dynamical $K/\pi$ fluctuations that are approximately independent of collision energy from $\sqrt{s_{NN}}$ = 7.7-200 GeV. This result is different than the observation of increasing fluctuations with decreasing incident energy observed by NA49. 
It is worth noting that the differences between the two experiments occur for dynamical ratio fluctuations involving kaons. There is good agreement between STAR and NA49 for $p/\pi$ dynamical fluctuations. A data point from RHIC at an energy below $\sqrt{s_{NN}}$ = 7.7 GeV would be provide additional information about the behavior of these particle ratio fluctuations at low energies.

\section{Summary}

Results from dynamical particle ratio fluctuations ($p/\pi$, $K/p$, and $K/\pi$) have been presented. 
Dynamical $p/\pi$ and $K/p$ fluctuations gradually increase from a larger negative value at $\sqrt{s_{NN}}$ = 7.7 GeV toward zero at $\sqrt{s_{NN}}$ = 200 GeV. Dynamical $K/\pi$ fluctuations are positive and approximately energy independent from $\sqrt{s_{NN}}$ = 7.7-200 GeV. There are differences between the measured values of $K/p$ and $K/\pi$ fluctuations at $\sqrt{s_{NN}}$ = 7.7 GeV from the STAR and NA49 experiments. Additional data at $\sqrt{s_{NN}}$ = 19.6 and 27 GeV will provide precision measurements of all fluctuation observables. Future runs at energies below $\sqrt{s_{NN}}$ = 7.7 GeV would be extremely valuable in determining the experimental trends of $p/\pi$, $K/p$, and $K/\pi$ fluctuations at these low energies and could help rectify some of the differences observed between the various experiments at SPS and RHIC.

\begin{acknowledgments}
We thank the RHIC Operations Group and RCF at BNL, the NERSC Center at LBNL and the Open Science Grid consortium for providing resources and support. This work was supported in part by the Offices of NP and HEP within the U.S. DOE Office of Science, the U.S. NSF, the Sloan Foundation, the DFG cluster of excellence `Origin and Structure of the Universe'of Germany, CNRS/IN2P3, FAPESP CNPq of Brazil, Ministry of Ed. and Sci. of the Russian Federation, NNSFC, CAS, MoST, and MoE of China, GA and MSMT of the Czech Republic, FOM and NWO of the Netherlands, DAE, DST, and CSIR of India, Polish Ministry of Sci. and Higher Ed., Korea Research Foundation, Ministry of Sci., Ed. and Sports of the Rep. Of Croatia, and RosAtom of Russia.
\end{acknowledgments}

\bigskip 

\end{document}